\documentclass[aps,floatfix]{revtex4-2}
\usepackage{dcolumn}
\usepackage{epsfig,epsf}
\usepackage{booktabs}
\usepackage{mathrsfs}
\usepackage{graphicx}
\usepackage{subcaption}
\usepackage{caption}
\usepackage{latexsym}
\usepackage{amsmath}
\usepackage{amssymb}
\usepackage{textcomp}
\usepackage{amsbsy}
\usepackage{tabularx}
\usepackage{rotating}
\usepackage{slashed}
\usepackage{array}
\usepackage{lipsum}
\usepackage[dvipsnames]{xcolor}
\usepackage{bm}
\usepackage{amsmath,amsfonts,amssymb}
\usepackage{bbold}
\usepackage{hyperref}
\usepackage{float}
\DeclareGraphicsExtensions{.png,.eps,.pdf}
\usepackage[left]{lineno}

\usepackage{color}
\usepackage[normalem]{ulem}

\definecolor{darkgreen}{cmyk}{1,0,1,0.4}

\long\def\/*#1*/{}

\usepackage[utf8]{inputenc}

\begin{document}

\title{Flavour from Fractal Mass Chains}
\author{Alejandro Ibarra}
\email{ibarra@tum.de}
\affiliation{Technical University of Munich, TUM School of Natural Sciences, Physics Department, 85748
Garching, Germany}
\author{Aadarsh Singh }
\email{aadarshsingh@iisc.ac.in}
\affiliation{Centre for High Energy Physics, Indian Institute of Science, C. V. Raman Avenue, Bengaluru 560 012, India }
\author{Sudhir K Vempati}
\email{vempati@iisc.ac.in}
\affiliation{Centre for High Energy Physics, Indian Institute of Science, C. V. Raman Avenue, Bengaluru 560 012, India }
\date{\today}

\begin{abstract}
We explore the possibility that the underlying flavour structure of the Standard Model could be determined by mass chains on a fractal geometry. We consider, as an example, the theory space on a Sierpinski-like geometry. The fermion mass chains on a Sierpinski-like geometry with three decorations (iterations) lead to three zero modes, which can be identified with the three generations of the Standard Model. This framework also reproduces the measured charged and neutral lepton masses and mixing angles with very few parameters. We also briefly discuss the possible extension to the quark sector.
\end{abstract}

\maketitle

\textbf{1. Introduction -} One of the major mysteries of the Standard Model of Particle Physics is the replication of the fundamental spin-1/2 fermions in three generations, with identical quantum numbers under the gauge symmetry $G_{\rm SM} = SU(3)_c\times SU(2)_L \times U(1)_Y$, but differing only in their mass. Such a successive repetition of generations in general implies complex Yukawa matrices, which lead to CP violation. In the quark sector, the Yukawa couplings take values from $\sim$ 1 for the top quark to $\sim$ $10^{-5}$ for the up/down quarks. The mixing angles between the generations vary from $\sin(\theta_{\rm mix}) \sim 0.22 - 10^{-3}$ over generations. In the leptonic sector, the mixing angles are sizable $\sim O(0.1)$ and at least one of the mass hierarchies is not as large as in the quark sector. 
Over the years several ideas have been proposed to address this puzzle, pioneered by the work of Froggatt and Nielsen \cite{froggatt1979hierarchy,froggatt1996fermion} to recent works on modular symmetry (some recent references include: \cite{dent2003modular,altarelli2006tri,king2007a4,feruglio2019neutrino,penedo2019lepton, Chen:2019ewa, Chen:2021prl,ding2024neutrino}) or quantum entanglement \cite{thaler2024flavor} (a summary of various approaches can be found in \cite{Babu:2009fd, Feruglio:2019ybq,feruglio2025quark,fernandez2025natural,altmannshofer2025recent}). 

In this work, we propose a novel approach to the flavour puzzle that simultaneously addresses the longstanding mystery of the existence of three generations. (For earlier work in this direction, please see \cite{kobayashi2010three,abe2008magnetized,abe2009magnetic,kan2024light,imai2024toward,abdalgabar2013evolution,libanov2001three,kaplan2001new,gogberashvili2007fermion,frere2001three,kaplan2012spacetime}.) We argue that the number of generations may originate from a replication property inherent in the theory space underlying the Standard Model. Such replication is characteristic of fractals, which are well-studied geometrical objects that naturally arise in various branches of physics, including nonlinear dynamics and complex systems \cite{malcai1997scaling}. To explore the possibility that the theory of flavour is connected to fractal structures, we will focus for concreteness on the Sierpinski triangle. This particular fractal has previously appeared in diverse contexts, ranging from biology, quantum materials, gravitation, to quantum computing \cite{sendker2024emergence, yang2022hofstadter,nottale1989fractals,canyellas2024topological}. Models of mass chains which are finite offer nice features like computability, renormalisability and thus testability. These fractal theory spaces were first considered in \cite{hill2003fractal}.

\textbf{2. flavour from the Sierpinski triangle -} 
We will consider deconstruction-like \cite{arkani2001electroweak,hill2001gauge, deBlas:2006fz}  mass chains \cite{giudice2017clockwork, Kaplan:2015fuy, Choi:2015fiu,craig2018exponential} on the Sierpinski geometry, with  Zero modes localised at different sites on this geometry. This leads to non-trivial patterns in their couplings to the Higgs. The Sierpinski triangle has been thoroughly studied as an archetype of a fractal, {\it i.e.} the repetition of a pattern on various scales and the creation of an intricate shape through repeated simple transformations. One starts with a very simple lattice (kernel lattice), which gives upon recurrent transformations a complex repeating pattern. After each step of transformation, the number of vertices increases to ($3\times n-3$) with $n$ being the number of nodes in the state being transformed. Using the notation of Graph Theory, the Hamiltonian for the kernel in the theory space can be written as
\begin{align}
    \mathcal{H}_{i,j} =a_i \delta_{i,j} + b_{ij}(1 - \delta_{i,j}) \;, \label{Hamiltonian}
\end{align}
with $i, j \in\{1,2,3\}$. The
Hamiltonian for the next iteration is constructed by replacing each $b_{ij}$ by $b_{i\alpha}$ and $b_{\alpha j}$, where $\alpha$ is the extra node introduced on the edge connecting two nodes. For each $(i,j)$ $\in$ $E_m$ and $i, j \in$ $V_m$ $\exists$ $(i,\alpha)$ and $(\alpha,j)$ $\in$ $E_{m+1}$ with $\alpha \in$ $V_{m+1}$ where $G_m$ = $(V_m, E_m)$ is the $m$-th recurrently generated graph, while $E_m$ and $V_m$ are edges and vertices respectively. Alternatively, one can use the node and edge labelling method described in \cite{jakovac2009vertex} to write a general form of Hamiltonian for $m$ recurrences. This notion can be straightforwardly applied to construct theory spaces with non-integer dimensions from dynamical graphs. 

The particle content consists of $n$ left- and right-handed fermions, with $n$ the number of vertices in the structure being considered. The corresponding Lagrangian for the mass chains reads: 
\begin{align} 
\mathcal{L}_{\rm S} = \mathcal{L}_{\rm kin} - \sum_{i,j=1}^{n} \overline{L_{i}}\mathcal{H}_{i,j}R_j + {\rm h.c.}  \label{Lagrangian}
\end{align}
where $\mathcal{H}_{i,j}$ is given in \eqref{Hamiltonian}.

\begin{figure}[t!]
    \includegraphics[scale=0.23]{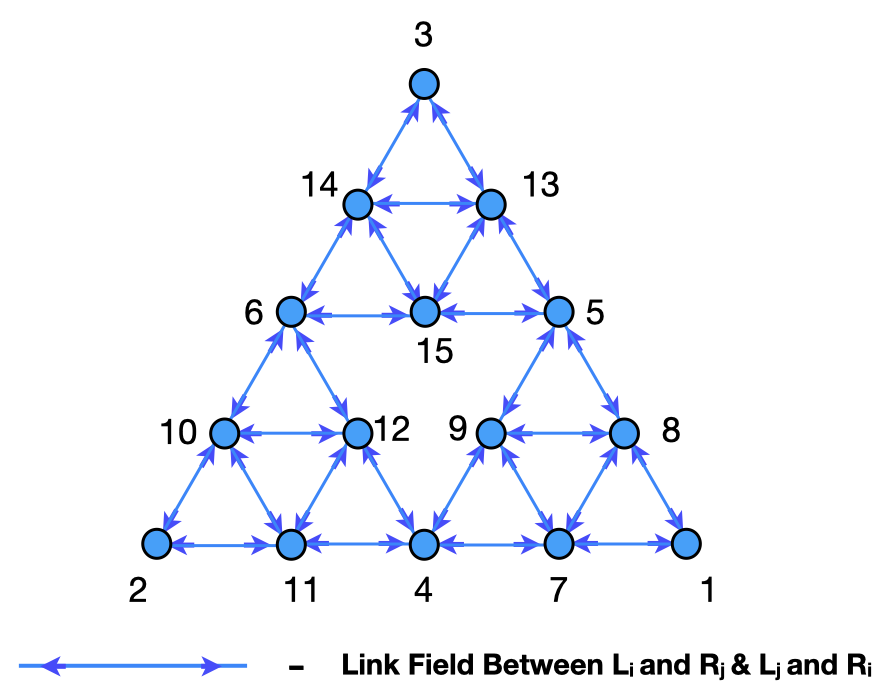}\\
    \caption{Fermions on the Sierpiński Fractal Graph.} \label{fractal}
\end{figure}

Starting from a single triangle, we consider the case with two further iterative transformations of the kernel lattice, which includes 15 left- and right-handed fermions. The corresponding figure is depicted in Fig. \ref{fractal}. The Lagrangian explicitly reads:
\begin{align}
  & \mathcal{L}_{\rm S} =  \mathcal{L}_{kin} -  \sum_{i,j=1}^{15} m_i \overline{L_{i}} \delta_{i,j}R_{j} + m \Bigl( \overline{L_{1}}q_{1,7}R_{7}+ \overline{L_{1}}   q_{1,8}R_{8}+ \overline{L_{7}}q_{7,4}R_{4}+ \overline{L_{7}}q_{7,9}R_{9}+ \overline{L_{7}}q_{7,8}R_{8}+ \overline{L_{8}}q_{8,5}R_{5}
  \nonumber \\ &
  + \overline{L_{8}}q_{8,9}R_{9}+ \overline{L_{4}}q_{4,9}R_{9} + \overline{L_{4}}q_{4,11}R_{11}+ \overline{L_{4}}q_{4,12}R_{12}+    \overline{L_{9}}q_{9,5}R_{5}+ \overline{L_{5}}q_{5,13}R_{13} + \overline{L_{5}}q_{5,15}R_{15}+ \overline{L_{2}}q_{2,10}R_{10} +  \nonumber \\ & \overline{L_{2}}q_{2,11}R_{11}+ \overline{L_{10}}q_{10,6}R_{6}+ \overline{L_{10}}q_{10,12}R_{12} +\overline{L_{10}}q_{10,11}R_{11}+  \overline{L_{11}}q_{11,12}R_{12}+ \overline{L_{6}}q_{6,12}R_{12} + \overline{L_{6}}q_{6,14}R_{14}+ \overline{L_{6}}q_{6,15}R_{15}  
  \nonumber \\ & + \overline{L_{3}}q_{3,13}R_{13}+ \overline{L_{3}}q_{3,14}R_{14}  + \overline{L_{3}}q_{3,15}R_{15}+ \overline{L_{13}}q_{13,14}R_{14}  +  \overline{L_{14}}q_{14,15}R_{15}  + (i\leftrightarrow j) \Bigr)  +{\rm h.c.}\;, \label{9} 
\end{align}
where $q_{i,j}$ are $\mathcal{O}(1)$ parameters as in a clockwork theory \cite{giudice2017clockwork}. The unusual labelling pattern follows the way the transformation acts on the original kernel.

The kinetic part of the Lagrangian has a symmetry
$U(15)_L \times U(15)_R$. This symmetry remains exact when $m_i=m$ for all $i$ and $q_{i,j}=0$ for all $i,j$, whereas it is completely broken when  $q_{i,j}$ are random. Remarkably, in the limit $q_{i,j} \neq 0$, there are three zero modes each for the left and the right fields, thus leading to a residual $U(3)_L \times U(3)_R$ symmetry. The zero modes need not be localised in general. However, for the particular choice $q_{i,j}$ = $f^{i-j}$ where $f \simeq O(1)$ and chosen positive, one has three zero modes which are all localised. The proof of this statement is presented in the Appendix \ref{Appendix-A}. 

\begin{figure}[t!]
    \includegraphics[width=0.5\textwidth]{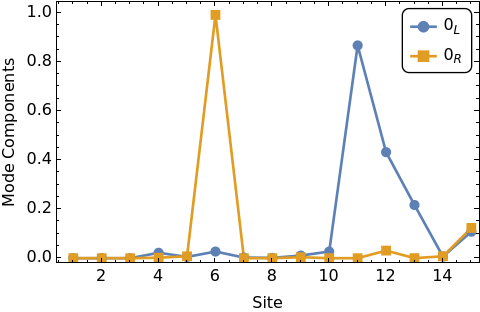}\\
    \caption{Left and Right 0-mode components on the Sierpiński Fractal Graph for f = 2.} \label{0-mode localization}
\end{figure}

This Lagrangian leads to a Dirac mass matrix of the following form, where we have assumed for simplicity $m_i$ = $2m$ and  $q_{i,j}$ = $f^{i-j}$:
\begin{align}
M_0 = m
\begin{pmatrix}
 2   &   f &   f^2 & 0 & 0 & 0 & 0 & 0 & 0 & 0 & 0 & 0 & 0 & 0 & 0 \\
   f^{-1} & 2   &   f &   f^2 &   f^3 & 0 & 0 & 0 & 0 & 0 & 0 & 0 & 0 & 0 & 0 \\
   f^{-2} &   f^{-1} & 2   & 0 &   f^2 &   f^3 & 0 & 0 & 0 & 0 & 0 & 0 & 0 & 0 & 0 \\
 0 &   f^{-2} & 0 & 2   &   f & 0 &   f^3 &   f^4 & 0 & 0 & 0 & 0 & 0 & 0 & 0 \\
 0 &   f^{-3} &   f^{-2} &   f^{-1} & 2   &   f & 0 & 0 & 0 & 0 & 0 & 0 & 0 & 0 & 0 \\
 0 & 0 &   f^{-3} & 0 &   f^{-1} & 2   & 0 & 0 &   f^3 &   f^4 & 0 & 0 & 0 & 0 & 0 \\
 0 & 0 & 0 &   f^{-3} & 0 & 0 & 2   &   f & 0 & 0 &   f^4 &   f^5 & 0 & 0 & 0 \\
 0 & 0 & 0 &   f^{-4} & 0 & 0 &   f^{-1} & 2   & 0 & 0 & 0 &   f^4 &   f^5 & 0 & 0 \\
 0 & 0 & 0 & 0 & 0 &   f^{-3} & 0 & 0 & 2   &   f & 0 & 0 & 0 &   f^5 &   f^6 \\
 0 & 0 & 0 & 0 & 0 &   f^{-4} & 0 & 0 &   f^{-1} & 2   & 0 & 0 &   f^3 &   f^4 & 0 \\
 0 & 0 & 0 & 0 & 0 & 0 &   f^{-4} & 0 & 0 & 0 & 2   &   f & 0 & 0 & 0 \\
 0 & 0 & 0 & 0 & 0 & 0 &   f^{-5} &   f^{-4} & 0 & 0 &   f^{-1} & 2   &   f & 0 & 0 \\
 0 & 0 & 0 & 0 & 0 & 0 & 0 &   f^{-5} & 0 &   f^{-3} & 0 &   f^{-1} & 2   &   f & 0 \\
 0 & 0 & 0 & 0 & 0 & 0 & 0 & 0 &   f^{-5} &   f^{-4} & 0 & 0 &   f^{-1} & 2   &   f \\
 0 & 0 & 0 & 0 & 0 & 0 & 0 & 0 &   f^{-6} & 0 & 0 & 0 & 0 &   f^{-1} & 2   \\
\end{pmatrix}\;.
\end{align}
As expected on general grounds, this mass matrix has three vanishing singular values. The null space is spanned by the left-handed and right-handed zero modes, and is given by 
\begin{align}
\Lambda_{i L} = 
 \begin{pmatrix}
 0 & f^{12} & -f^{11} & 0   & -f^9 & 2 f^8 & 0 & 0    & -f^5 & -f^4  & 0 & 0 &0 & 1 &0\\
 0 & 0      & 0       & f^9 & -f^8 & f^7   & 0 & -f^5 & 0    & -f^3  &0 & 0 & 1 & 0 &0 \\
 0 & -f^{10} & f^9 & 2 f^8 & -f^7 & 0 & -f^5 & -f^4 & 0 & 0 & 0 & 1 & 0 &0 &0 
\end{pmatrix} \;,\label{left-0 mode}
\end{align} 
\begin{align}
 \Lambda_{i R} =
\begin{pmatrix} 0 & {f^{-12}} & -{f^{-11}} & 0 & -{f^{-9}} & 2{f^{-8}} & 0 & 0 & -{f^{-5}} & -f^-{4} & 0 & 0 & 0 & 1 & 0 \\
 0 & 0 & 0 & f^{-9} & -f^{-8} & f^{-7} & 0 & -f^{-5} & 0 & -f^{-3} & 0 & 0 & 1 & 0 & 0 \\
 0 & -{f^{-10}} & f^{-9} & 2f^{-8} & -f^{-7} & 0 & -f^{-5} & -f^{-4} & 0 & 0 & 0 & 1 & 0 & 0 & 0
 \end{pmatrix}\;, \label{right-0 mode}
\end{align} 
with each row representing one eigenvector. The localisation of left and right-handed zero modes is evident from figure \ref{0-mode localization}. As can be seen from eq.\eqref{left-0 mode} and eq.\eqref{right-0 mode}, the modes are delocalised for $|f| =1$ and localized for any other value of $f$.

In our scenario, the left chiral zero modes of L are identified with the three generations of SM lepton doublets, whereas the right chiral zero modes of $E$ and $N$ are identified with the three generations of SM lepton singlets and right-handed neutrinos. We propagate the $L^{e}, E, N$ fields on the fractal geometry \footnote{{\color{black} The unphysical chiral zero modes can form the dark sector, which can be made heavy with a dark Higgs. Another possibility is to use a discrete symmetry to project them out
as happens in extra dimension theories with orbifold \cite{perez2005introduction,csaki2016beyond}.}}. Assuming that the Higgs field couples with equal strength at all sites, the Higgs interaction Lagrangian can be written as:
\begin{align}
\mathcal{L}_{\mathcal{H}}=-Y_{ij}^{\nu}  \bar{L}_i^{e} \widetilde{H} N_j -Y_{ij}^e  \bar{L}_i^{e} H E_j+\text { h.c. } \hspace{1cm} i,j = \{1,2,3,...,15 \} \label{interaction} 
\end{align}
where $L^e$, $E$, and $N$  are respectively the left-handed lepton and right-handed electron and neutrino fields, while $Y^{\nu}$ and $Y^{e}$ are the $15 \times 15$ neutrino and charged lepton Yukawa couplings. After the Higgs attains a vev, the total Lagrangian has the form (from eq.(\ref{Lagrangian}) and eq.(\ref{interaction}) )
\begin{equation}
\mathcal{L}_{\text {Tot }} = \mathcal{L}_{\rm Kin} -\sum_{i,j=1}^{n} \overline{L_{i}^{e}}\mathcal{H}_{i,j}^eE_j - \sum_{i,j=1}^{n} \overline{L_{i}^{e}}\mathcal{H}_{i,j}^{\nu}N_j + \mathcal{L}_{\mathcal{H}} + {\rm h.c.}
\end{equation}
Diagonalising $\mathcal{H}^e$ and $\mathcal{H}^{\nu}$, and redefining fields as $L_{\nu}^{e}$ = $\mathcal{U}^{\nu} \chi_{L}^{\nu}$, $L_{e}^{e}$ = $\mathcal{U}^{e} \chi_{L}^e$, $N$ = $\mathcal{V}^{\nu} \chi_{N}^{\nu}$ and $E$ = $\mathcal{V}^{e} \chi_{E}^{e}$, with $U$ unitary matrices, the interaction terms of the Lagrangian become 
\begin{equation}
\mathcal{L}_{\mathcal{H}} \supset -Y^{\nu} \bar{\chi}_{L}^{\nu} \mathcal{U}^{\nu \dagger} \widetilde{H} \mathcal{V}^{\nu} \chi_{N}^{\nu} -Y^{e} \bar{\chi}_{L}^{e} \mathcal{U}^{e \dagger} \widetilde{H} \mathcal{V}^{e} \chi_{E}^{e}  + {\rm h.c.}
\end{equation}

Lastly, integrating out the heavier modes, one obtains the effective $3 \times 3$ Yukawa couplings for the three Standard Model generations. Considering the corrections from the heavy modes to be small $\sim O(v/m)$, one can use the Moore-Penrose pseudoinverse \cite{dresden1920fourteenth,bjerhammar1951rectangular,penrose1955generalized} to derive the $3 \times 3$ Yukawa matrix for the three generations. This has the form
\begin{align}
\frac{Y_{yuk}^{\nu}}{Y} \approx \scalebox{0.6}{$
\begin{bmatrix}
\scalebox{1.1}{$f_L^{-12} f_N^{-12} \left(
f_L^{12} + f_L^{11} f_N + f_L^9 f_N^3 + 4 f_L^8 f_N^4 + f_L^5 f_N^7 + f_L^4 f_N^8 + f_N^{12}
\right)$}
 & \scalebox{1.4}{$f_L^{-8} f_N^{-3}+2f_L^{-4} f_N^{-7}+f_L^{-3} f_N^{-8}$} & {\scalebox{1.4}{$-f_L^{-1} f_N^{-9}-f_N^{-10}+f_L^{-3} f_N^{-7}$}} \\[0.5em] 
\scalebox{1.4}{$f_L^{-6} f_N^{-4}+2f_L^{-2} f_N^{-8}+f_L^{-1} f_N^{-9}$} & \scalebox{1.1}{$\left( f_L^9 + f_L^8 f_N + f_L^7 f_N^2 + f_L^5 f_N^4 + f_L^3 f_N^6 + f_N^9 \right) f_L^{-9} f_N^{-9} $}
 & \scalebox{1.4}{$f_L^{-4} f_N^{-4}+f_L^{-1} f_N^{-7}+2f_N^{-8}$} \\[0.5em] 
\scalebox{1.4}{$f_L^{-7}+f_L^{-3} f_N^{-4}+f_L^{-2} f_N^{-5}$} & \scalebox{1.4}{$-2f_L^{-6}-f_L^{-3} f_N^{-3}-f_L^{-1} f_N^{-5}$} &  \scalebox{1.4}{$f_L^{-5}-f_L^{-1} f_N^{-4}+f_N^{-5}$} \\[0.5em] 
\end{bmatrix}.
$}
\end{align}
The charged lepton Yukawa matrix can be derived in a similar way, leading to an identical results, albeit with different parameters, $f_i$ and $Y^e$. 

The number of free parameters can be further reduced by choosing the Higgs localization suitably at various sites. One of the simplest choices to localise the Higgs is at 4, 9 and 13 sites \footnote{This can be achieved if: (a) the Higgs field is localised on three different sites akin to the extra-dimensional picture \cite{arkani2000hierarchies,abe2018atmospheric} or (b) there are three Higgs fields localised at different sites as in three Higgs doublet models \cite{branco1985resuscitation, FLORES198395}.}, which gives the Lagrangian to be
\begin{equation}
\mathcal{L}_{\mathcal{H}}=-y_1^{\nu}  \bar{L}_{4}^e \widetilde{H} N_{4}-y_2^{\nu}  \bar{L}_{9}^e \widetilde{H} N_{9}- y_3^{\nu}  \bar{L}_{13}^e \widetilde{H} N_{13}-y_1^e  \bar{L}_{4}^e H E_{4}-y_2^e  \bar{L}_{9}^e H E_{9}- y_3^e  \bar{L}_{13}^e H E_{13}
+\text { h.c. } \label{eq.int}
\end{equation} 
Assuming for simplicity, $y_1^e \simeq y_2^e \simeq y_3^e \simeq y_1^{\nu} \simeq y_2^{\nu} \simeq y_3^{\nu} = Y$  for both charged and neutral lepton mass matrices and $f_i\gg$ 1, the PMNS matrix reads:
\begin{align}
U_{\rm PMNS} \approx  
\scalebox{0.62}{$
\begin{pmatrix}
f_L^{-2} f_E^{-4} f_N^{-4} + f_L^{-4} (2 f_L + f_E)^{-1} (2 f_L + f_N)^{-1} f_E f_N \scalebox{1.2}{$+ 1$}
& 
\scalebox{1.2}{$
2 (f_E - f_N) \, f_L^{-1} (2 f_L + f_E)^{-1} (2 f_L + f_N)^{-1}$}
& 
\scalebox{1.0}{$
\left( f_E^{-4} - f_E \, f_N^{-4} (2 f_L + f_E)^{-1} \right) f_L^{-1} - 2 \left( f_N^4 (2 f_L + f_N) \right)^{-1}$} 
\\[0.5em]
\scalebox{1.2}{$f_N (2 f_L^3 + f_L^2 f_N)^{-1} - f_E (2 f_L^3 + f_L^2 f_E)^{-1}$}
& 
\scalebox{1.2}{$f_E f_N (2 f_L^3 + f_L^2 f_E)^{-1} (2 f_L^3 + f_L^2 f_N)^{-1} + 1$}
& 
\scalebox{1.3}{$f_L (f_E^{-4} - f_N^{-4})$}
\\[0.5em]
\scalebox{1}{$- f_N f_E^{-4} (2 f_L^2  + f_L f_N)^{-1} - 2 (2 f_L f_E^4 + f_E^5)^{-1} + f_L^{-1} f_N^{-4}$}
& 
\scalebox{1.3}{$f_L (f_N^{-4} - f_E^{-4})$}
& 
\scalebox{1.3}{$f_L^2 f_E^{-4} f_N^{-4} + 1$}
\end{pmatrix}\label{pmns}
$}
\end{align}

\begin{figure}[t!]
    \vspace{0.45cm}
    \includegraphics[width=0.5\textwidth]{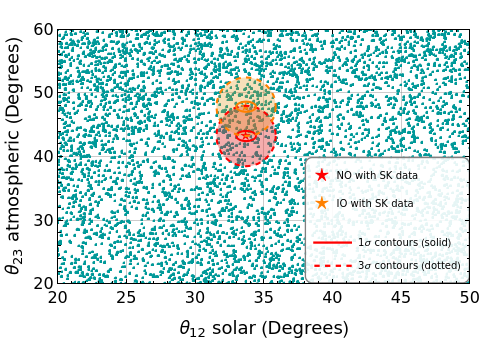}
    \caption{Leptonic mixing angles from a random scan of the parameters of the model (for details, see the main text).  The contours and stars show respectively the preferred region and best fit values for the solar $\theta_{12}$ and atmospheric $\theta_{23}$ mixing angles with SK data for NO (red) and IO (orange) \cite{esteban2025nufit}.} \label{PMNS_Angles}
\end{figure}

\begin{figure}
    \includegraphics[width=0.32\textwidth]{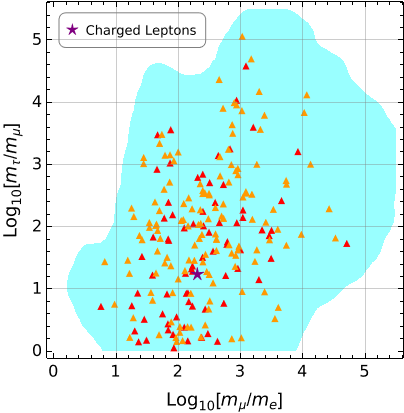}
    \includegraphics[width=0.32\textwidth]{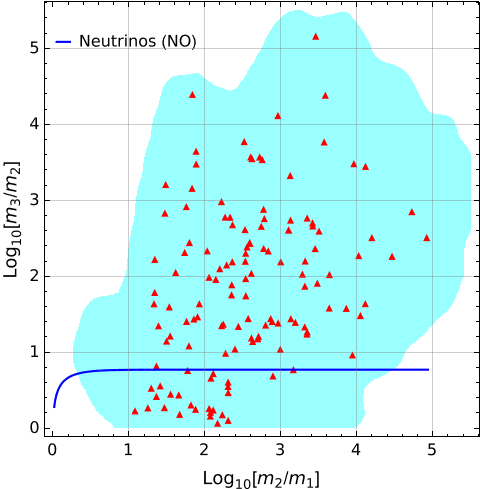}
    \includegraphics[width=0.32\textwidth]{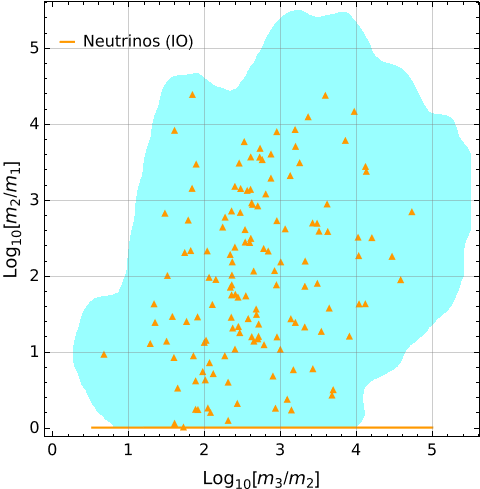}
    \caption{Mass hierarchies in the charged lepton sector (left panel) and in the neutrino sector (middle and right panel) from a random scan of the parameters of the model. The orange triangles indicate points in the scan leading to mixing angles within a 3-sigma range of observations with Inverted Ordering(IO), and red triangles for 3-sigma with Normal Ordering(NO) (shaded region in Fig.~\ref{PMNS_Angles} in respective colour). The purple star indicates the measured charged lepton mass hierarchies \cite{navas2024review}, while the blue line (middle plot) and orange line (right plot) indicate the neutrino mass hierarchies compatible with oscillation experiments for Normal Ordering and Inverted Ordering, respectively \cite{esteban2025nufit}. Here, $m_3>m_2>m_1$ for both charged leptons (left) and NO neutrinos (middle). For the IO neutrinos (right), the convention is $m_2>m_1>m_3$.}     \label{mass_hierarchy}
\end{figure}

Fig.~\ref{PMNS_Angles} shows a scatter plot of mixing angles for a random scan of the model parameters $f_L$, $f_E$, $f_N$, $y_{e1}$, $y_{e2}$, $y_{e3}$, $y_{\nu 1}$, $y_{\nu 2}$, $y_{\nu 3}$ within the range [0.1, 10] and $m = 1$ TeV.  The plot also shows the experimental values of the mixing parameters for the normal and inverted mass orderings \cite{esteban2025nufit}. In Fig.~\ref{PMNS_Angles}, the shaded red and orange region corresponds to the 3-sigma regions for Normal and Inverted mass hierarchies from experimental data (with SK atmospheric data), respectively. Clearly, the model can accommodate sizable leptonic mixing angles. Fig.~\ref{mass_hierarchy} shows the mass hierarchies between the heaviest $m_3$ and the next-to-heaviest generation $m_2$, and between the next-to-lightest $m_2$ and the lightest generation $m_1$, for the same random scan as in Fig.~\ref{PMNS_Angles}, for the charged leptons (left panel) as well as for the neutrinos (middle and right panel). The cyan region corresponds to the ranges of mass ratios obtained in our scan, while the orange and red triangular points correspond to points that reproduce the leptonic mixing angles (points that lie within the shaded disk in Fig.~\ref{PMNS_Angles}) for IO and NO, respectively. The model tends to generate large fermion hierarchies, thus easily accommodating the observed charged lepton hierarchies (indicated as a purple star in Fig.~\ref{mass_hierarchy} (left)), and the observed neutrino mass hierarchies (indicated as a blue and orange line in Fig.~\ref{mass_hierarchy} for the normal mass hierarchies (middle) and Inverted mass hierarchies (right)). Obtaining the correct neutrino masses generally requires larger values of $f$. In this limit, the fractal construction naturally generates large mass hierarchies, as reflected in the scattered plot; nevertheless, viable parameter sets that match the observed neutrino masses can still be found. The mechanism does prefer producing larger mass hierarchies compared to smaller ones, like for the neutrino masses. However, there is still good room for the neutrino mass and mixing. Stabilisation of the fields is good, as we have checked explicitly. Our scenario assumes Dirac neutrino masses, although the framework could also be extended to include Majorana masses.


This framework could also be extended to the mixing and masses in the quark sector of the SM. We found for parameter $f_i={\cal O}(1)$ with $m = 1$ TeV, this fractal generates quark masses and mixing angles in the ballpark of the measured values. For instance, for $\{ y_1, y_2, y_3 \}$ = $\{0.07, 0.04, 0.001\}$, and $f_Q = 1.2$ and $f_D$ = 1.9, the resulting down type masses generated are $O(7)$ MeV, $O(0.1)$ GeV, $O(4)$ GeV; for up-sector masses, with $\{ y'_1, y'_2, y'_3 \}$ = $\{3.7, 0.002, 0.06 \}$ and $f_U$ = 1.4, the up-type quark masses generated are $O(8)$ MeV,  $O(1.2)$ GeV and $O(172)$ GeV. See Appendix~\ref{Appendix-B} for the full spectrum of mass eigenvalues obtained from the fractal model, listed in Table~\ref{tab:mass-values-indexed} for $f$ = 1.2.

\textbf{3. Conclusions $\&$ Outlook - }  
Fractal geometry is widely observed in Nature, where intricate patterns and self-similarity emerge across multiple scales.\cite{malcai1997scaling}. In this letter, we have constructed a theory of flavour based on the mass chains on fractal geometry. For concreteness, we have chosen the Sierpinski Triangle with up to three iterations (the first iteration being a dot). This naturally leads to three zero modes, which we identify with the three generations of the Standard Model.  
The mass matrices for neutrinos and charged leptons can be parameterised in terms of a few parameters (up to $\mathcal{O}(1)$ Yukawa couplings) and can accommodate the measured leptonic masses and mixing angles for suitable choices of the parameters. Analogous conclusions can be obtained for the quark sector. The model is renormalizable, and the mode is stable under quantum corrections. The new vector fermions are heavy with small couplings to the Higgs and might lead to interesting phenomenology. It would be interesting to explore whether this fractal theory can be incorporated into a UV-complete model or whether other fractal structures can also reproduce the measured parameters.

\paragraph*{\bf Acknowledgements:}
SKV is supported by SERB, DST, Govt. of India Grants MTR/2022/000255 , “Theoretical aspects of some physics beyond standard models”, CRG/2021/007170 “Tiny Effects from Heavy New Physics “and IoE funds from IISC. AS thanks CSIR, Govt. of India for SRF fellowship No. 09/0079(15487)/2022-EMR-I. AI is supported by the Collaborative Research Center SFB1258 and by the Deutsche Forschungsgemeinschaft (DFG, German Research Foundation) under Germany's Excellence Strategy - EXC-2094 - 390783311.

\appendix

\section{Mathematical results} \label{Appendix-A}

Consider any matrix A with a non-zero kernel space dimension, and elements $a_{i,j}$. We define the matrix $B$ with elements as,
\begin{align}
b_{i,j} = \frac{a_{i,j}}{f^{(i-j)}}, \hspace{2cm} \forall \hspace{.2cm} f \in \mathbb{R} \setminus \{0\}.
\label{eq:B&A}
\end{align}
Then, the following results hold for the matrix $B$.

\textbf{Corollary 1} - The rank of the matrix B is equal to the rank of A. Namely, the original rank-nullity of A is preserved by B.

\textbf{Proof} -  We denote $v_{i_0}$ ($v'_{i_0}$) as the $i_0$-th row of the matrix A (B) of dimensions $N \times M$. The nullity in A implies that $v_{i_0}$ is linearly  dependent on other rows {\it i.e.},
\begin{align} 
v_{i_0}  = \sum_{j \neq i_0}^N \alpha_j v_j.
\label{eq:nullity-A}
\end{align}
Take the $k^{th}$ element of $i_0$,
\begin{align}
v_{i_0,k} = \sum_{j \neq i_0}^N \alpha_j v_{j,k},
\end{align}
where $\alpha_j$ is the same for a given row. 

Then from the definition of elements of matrix B ({\it cf.} Eq.~(\ref{eq:B&A}))
\begin{align}
v'_{i_0,k} &= \frac{v_{i_0,k}}{f^{(i_0-k)}} 
\end{align}
and replacing in Eq.~(\ref{eq:nullity-A}), one obtains
\begin{align}
v'_{i_0}  = \sum_{j \neq i_0}^N \alpha'_j v'_j,
\end{align}
with $\alpha'_j$ = $\alpha_j f^{(j-i_0)} $. Therefore, nullity in A will lead to nullity in B. Similarly, nullity in matrix B will lead to nullity in A. A generalised version of this Corollary can be found in \cite{singh2024certain}.

\textbf{Corollary 2} - For any matrix A with $\{v^1, v^2,\hdots,v^n\}$ as eigenvectors of its nullspace, the corresponding eigenvectors for the nullspace of matrix B are given by $\{v'^1, v'^2,\hdots,v'^n\}$ with
\begin{align} v'^i_j = v^i_jf^{(-j)}, \hspace{2cm} \forall \hspace{.2cm} f \in \mathbb{R} \setminus \{0\}
\end{align}
$v^i_j$ represents the $j^{th}$ component of $i^{th}$ null basis vector.

\textbf{Proof} - Consider the $v^i$-th null basis vector of matrix A, $Av^i = \Vec{0}$. This implies
\begin{align}
\sum_{j=1}^{M} a_{l,j}v^i_j = 0 \hspace{1cm} \forall \hspace{.2cm} l \in \{1,2,\hdots,N\} ,
\end{align}
now using the element-wise transformation of matrix A by the operator in the above corollary, $a_{l,j} = b_{l,j} \times f^{(l-j)}$, it follows that
\begin{align}
\sum_{j=1}^{M} b_{l,j}\times f^{(l-j)}v^i_j = 0 \hspace{1cm} \forall \hspace{.2cm} l \in \{1,2,\hdots,N\} .
\end{align}
Finally, defining $v'^i_j = v^i_j f^{-j}$ and simplifying one obtains:
\begin{align}
&\sum_{j=1}^{M} b_{l,j}v'^i_j = 0 \hspace{1cm} \forall \hspace{.2cm} l \in \{1,2,\hdots,N\}  .
\end{align}
Hence, all of the null basis vectors of A with their elements scaled to some power of $f$, will behave as null basis vectors for matrix B.

\section{Mass Eigenvalues} \label{Appendix-B}
This appendix presents the masses of Vector like particles for a field on the fractal geometry, the non-zero modes, ordered from largest to smallest. The eigenvalues  before the Higgs SSB, are given by $\lambda = \{5,77 m$, $4.96 m$, $4.96 m$, $2.84 m$, $2.84 m$, $2.71 m$, $1.74 m$, $1.74 m$, $ m$, $0.51 m$, $0.44 m$, $0.44 m$, $0$, $0$, $0\}$. The values in Table~\ref{tab:mass-values-indexed} correspond to the numerical results discussed in Section 2, for m = 1 TeV. This Table also shows the couplings generated for one of the three left-handed 0-modes $0_L$ with the SM Higgs and right-handed heavy modes $\chi_{R,i}$ for $y_1$ = $y_2$ = $y_3$ = 1 and $f$ = 1.2.

\begin{table}[ht]
\centering
\begin{tabular}{|c|c|c|}
\hline
\textbf{Index} & \textbf{Mass (TeV)} & \textbf{Higgs Coupling} \\
\hline
1 & 6.72 & 0.0101 \\
\hline
2 & 6.03  & 0.0102  \\
\hline
3 & 5.48 & 0.0241 \\
\hline
4 & 3.63 & -0.0132 \\
\hline
5 & 3.38 & -0.00127 \\
\hline
6 & 3.09 & -0.00935 \\
\hline
7 & 2.10 & 0.03 \\
\hline
8 & 1.97 & 0.0404  \\
\hline
9 & 1.33 & 0.027  \\
\hline
10 & 0.575 & -0.00299 \\
\hline
11 & 0.407 & -0.0202  \\
\hline
12 & 0.379 & -0.0288  \\
\hline
\end{tabular}
\caption{Mass eigenvalues and corresponding couplings for the fractal model.}
\label{tab:mass-values-indexed}
\end{table}

\bibliographystyle{unsrt}

\bibliography{bib}

\end{document}